# An algorithm for determining the state of a non-stationary dynamic system for assessing fire safety control in an enterprise by the method of integrated indicators

S N Masaev[1], A N Minkin and D A Edimichev

Institute of Oil and Gas, Siberian Federal University, pr. Svobodnyj, 79, Krasnoyarsk, 660041, Russia

[1] E-mail: faberi@list.ru

**Abstract**. Analysis of the scientific literature showed that a lot of work is devoted to assessing the effectiveness of fire safety management in an enterprise. It is worth noting that today there is no universal method for the integrated assessment of fire safety management, taking into account the interconnectedness of all enterprise subsystems and the influence of environmental factors. One of the original approaches to assessing the effectiveness of the fire safety management system is the method of integral indicators. The method of integral indicators is used in the algorithm for analyzing the state of a dynamic non-stationary system for assessing fire safety management in an enterprise. The algorithm is implemented in the author's complex of programs described in the text of the article. In the simulation, an analysis of 1.2 million values is performed on a well-studied economic object with the spaces identified at each time step: actual data, control and environmental parameters. In the experiment, the basic mode of operation of the enterprise does not contain the implementation of a fire safety management strategy. The research showed significant changes in the values of the integral indicator characterizing the state of the enterprise during the implementation of the fire safety management system at the enterprise.

## 1. Introduction

In the theory of management, the activity of an enterprise, as a system, is studied in various ways: intersectoral balances, numerical, vector, parametric and neural network modeling, agent approach, etc. however, the fire safety control loop is not considered systematically. One of the reasons is the difficulty of identifying an economic object as a system when performing functional modes with various restrictions and the influence of environmental parameters.

Control issues involved: V. V. Leontiev and L. V. Kantorovich, A. G. Granberg, A. G. Aganbegyan, V. F. Krotov [1-8]. The problems of dynamical systems were dealt by other authors of [9-17]. In 2009, the author proposed integral indicators [18] to determine the financial crisis of 2008 based on the method of correlation adaptometry [19]. In the future, the method of integrated indicators was used at the enterprises to evaluate: in 2014, personnel management strategies [20], in 2016, quality management strategies (TQM) [21], in 2018, PMBOK project management [22], in 2019. Management of special economic zones [23-25], in 2019 determining the damage from sanctions [26, 27].

The aim of this work is: to assess the state of the enterprise with an integrated indicator as a multidimensional dynamic system, when managing fire safety at the enterprise. To achieve the goal, you must complete the following tasks:
- Formalize an economic object as a multidimensional dynamic system (hereinafter the system);
- In the system, assign to the functions to be performed a list of the corresponding business processes of the enterprise (basic strategy).
- Set the execution mode for a set of business processes, taking into account the implementation of the fire safety management strategy at the enterprise.
- Assess the integral state of the parameters characterizing the basic strategy and strategy of fire safety management at the enterprise.

## 2. Method

First, the activities of the enterprise are identified as a system $S=\{T,X\}$, where $T=\{t: t=1,...,T_{max}\}$ are many points in time; $X$ - space of system parameters; $x(t)=[x^1(t), x^2(t),...,x^n(t)]^T \in X$ – $n$ – vector of values corresponding to the state of the system. The values of the vector $xi(t)$ - the value of financial expenses and income of the enterprise. The dimension of the system $n$ is 1.2 million parameters. Based on the parameters $X$ and $T$, we consider the enterprise a multidimensional dynamic system (hereinafter referred to as the system).

The analysis of the system at the moment $t$ is performed $x(t)$ on the basis $k$ of previous measures. The parameter $k$ is the length of the time series segment (accepted $k=6$ months in the work). Then we have a matrix:

$$X_k(t) = \begin{bmatrix} x^T(t-1) & x^T(t-2) & ... & x^T(t-k) \end{bmatrix}^T \qquad (1)$$

The value of the time series can be used to identify the frequency of the influence of external factors on the system. The issue was considered separately in [28].

Following the method of integral indicators [18], we calculate the mutual correlation coefficients between the values $x(t)$ of the parameters characterizing the state of the system for the entire period of economic activity of an economic object (enterprise). We get the correlation matrix $R_k(t)$:

$$R_k(t) = \frac{1}{k-1} \overset{o}{X}_k^T(t) \overset{o}{X}_k(t) = \|r_{ij}(t)\|, \qquad (2)$$

$$r_{ij}(t) = \frac{1}{k-1} \sum_{l=1}^{k} \overset{o}{x^i}(t-l) \overset{o}{x^j}(t-l), \; i,j = 1,...,n, \qquad (3)$$

where $t$ are the time instants, $r_{ij}(t)$ are the correlation coefficients of the variables $x_i(t)$ и $x_j(t)$ at the time instant $t$.

Next we form one of the four integral indicators, the sum of the absolute values of the correlation coefficients, an indicator for express estimation of the correlation of system parameters $G_i(t)$:

$$R_i(t) = G_i(t) = \sum_{j=1}^{n} |r_{ij}(t)|. \qquad (4)$$

The state of the entire system is calculated as:

$$G = \sum_{t=1}^{T=\max} \sum_{i=1}^{n} G_i(t) \qquad (5)$$

The implementation of the method is performed in the author's complex of programs.

## 3. Author's software package
The author's software package consists of four separate programs: RosPatent Certificate on registration of a computer programs: 1. No. 2013614410, 2. No. 2017616973, 3. No. 2008610295, 4. No. 2017616970). This software package is used to perform the experiment algorithm.

## 4. Experiment algorithm
The step number in the algorithm coincides with the number of the program being executed from the author's program complex described above.

1 step. We load data on the actual activity of the economic object $x^i(t)$, management data for past periods of time and data characterizing the external environment. We identify the enterprise as a multidimensional dynamic system. We introduce a fire safety management strategy at the enterprise. We calculate the integral indicator for all states of the system. We set the objective function. If the data characterize the object, then go to step 2, otherwise repeat 1 step.

2 step. In the loaded model, we select the form of representation of the control method. For example, PMBOK, budgeting, Cobb-Douglas formula, V-shaped project management model, quality management system (TQM), life cycle assessment method (products, enterprises, strategies), etc. Additional examples can be found in separate works [20-27]. In our case, the fire safety requirements affect the control of system conditions, affecting all other parameters $x^i(t)$ (1). We check the completeness of the system description with the selected technique. If the syntax of the control method satisfies our requirements, then go to step 3, otherwise we return to the selection of a new control method step 2. If the method for control is not found, but the description and control of the system only by the name of the variables $x^i(t)$ satisfies us, then go to step 3, otherwise 1 step.

3 step. We check the control for optimality [29]. If the solution is not optimal, according to the set objective functions, and does not suit us, then we return to step 1, otherwise we proceed to step 4.

4 step. We evaluate the effectiveness of the fire safety management system. If efficiency does not suit us, then we set new control actions and go to step 1, otherwise the end of the algorithm.

The algorithm is performed according to the parameters of the object of study.

## 5. Characteristics of the research object
The enterprise used as an example has a deep wood processing plant with non-waste production. In order to assign the developed model the status of a priority investment project of the Russian Federation, this enterprise passed the expertise of the Ministry of Natural Resources of the Krasnoyarsk Territory, Rosleskhoz, the Ministry of Industry and Trade, the Ministry of Economy and Development of the Russian Federation [30]. After 1.5 years, production began. The project is implemented at the expense of the bank's credit and the use of tax benefits for priority projects.

The functions of the fire safety department also include the development of fire safety documentation [31-34].

## 6. Experiment results
Basic strategy $G_1^6 =1\ 229\ 156$ is figure 1, characterizes the activities of the enterprise without a fire safety management system. Strategy 2 $G_2^6 =1\ 248\ 571$, characterizes the existing fire safety management system at the enterprise is figure 1.

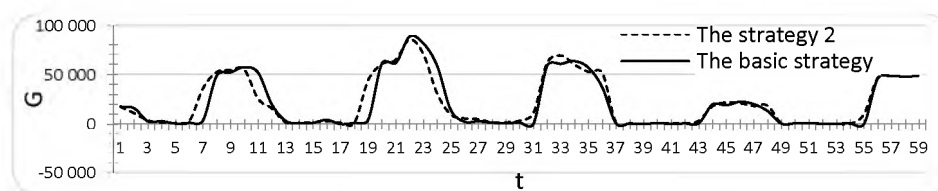

Figure 1. Dynamics of the value for two simulated options for the development of the company.

Then the influence of the fire safety management system is measured as the difference between the state of the control system parameters $\Delta G = G_1^6 - G_2^6$ and is $\Delta G$ equal to minus 19 415. The value $G_2^6 > G_1^6$ is due to the higher multifactorial nature of the system (parameter interconnectedness) with the influence of environmental parameters.

## 7. The discussion of the results

When the company's turnover in revenue for 5 years is 10.451 billion rubles. The fire safety management system is characterized by the following economic parameters: the costs of the fire safety system 0.283 billion rubles, the prevented losses of the enterprise from the fire are 6 billion rubles, the payback period is 1.5 months.

## 8. Conclusion

The tasks set at the beginning of the work were completed:

- The economic object (enterprise) is set as a multidimensional dynamic system $S$ according to the parameters of spaces $X$ and $T$.
- Assigned to the performed functions of the system $x_i$, characterizing the basic strategy of the enterprise $G_k$.
- The mode of execution of a set of business processes was set taking into account the implementation of the fire safety management strategy at the enterprise.
- An assessment is made by the method of integral indicators of two states of a non-stationary dynamic system: the basic strategy $G_1^6$ - 153 080 and strategy 2 $G_2^6$ - 155 896.

The goal set at the beginning of the work, to assess the state of the enterprise with an integrated indicator as a multidimensional dynamic system, has been achieved in managing fire safety at the enterprise.